# Electron and Phonon Temperature Relaxation in Semiconductors Excited by Thermal Pulse.

Yu. Gurevich[1], G. N. Logvinov[1], A. F. Carballo-Sanchez[2],

Yu. V. Drogobitskiy[3], J. L. Salazar[1]

[1] Depto. de Física, CINVESTAV-I.P.N. Apdo. Postal 14-740, México, 07000 D.F. México

[2] Instituto de Física, Universidad Autónoma de Puebla, Apdo. Postal J- 48, Puebla, Pue. 72570, México

[3] Ternopil Pedagogical State University, Krivonosa str., 2, P. O. 46027, Ternopil, Ukraine

E-mail: gurevich@fis.cinvestav.mx

## Abstract

Electron and phonon transient temperatures are analyzed in the case of nondegenerate semiconductors. An analytical solution is obtained for rectangular laser pulse absorption. It is shown that thermal diffusion is the main energy relaxation mechanism in the phonon subsystem. The mechanism depends on the correlation between the sample length $l$ and the electron cooling length $l_{\varepsilon}$ in an electron subsystem. Energy relaxation occurs by means of the electron thermal diffusion in thin samples ( $l << l_{\varepsilon}$ ), and by means of the electron-phonon energy interaction in thick samples ( $l >> l_{\varepsilon}$ ). Characteristic relaxation times are obtained for all the cases, and analysis of these times is made. Electron and phonon temperature distributions in short and long samples are qualitatively and quantitatively analyzed for different correlations between the laser pulse duration and characteristic times.

**I. Introduction**

The study of thermal processes and properties of semiconductors has always been and still is one of the main problems of solid state physics. This topic is being intensively developed at present due to the successful application of different laser pulse techniques. Some of them belong to excitation and probe techniques, such as time resolution of resolved refractivity, optical absorption, Raman scattering, luminescence and others techniques described in Ref. [1]. Other methods involving transient thermoelectric responses are discussed in [2, 3].

The development of methods that provided new experimental data stimulated the development of an adequate theory of transient thermal processes in semiconductors. Some of these results have been published in Refs [4] and [5]. The former paper describes insulators and the latter describes unipolar nondegenerate semiconductors where a two-temperature model was employed. Within the framework of this model, the separate nonequilibrium temperatures describe the energy nonequilibrium states of electrons and phonons. These temperatures are produced in accordance with an electron-phonon energy interaction and separate thermal boundary conditions (see e.g. [6]).

In [5] $\kappa_p$ and $\alpha_e$ were neglected for simplicity since in nondegenerate semiconductors the electron thermal conductivity $\kappa_e$ is essentially less than the phonon thermal conductivity $\kappa_p$, and the phonon thermal diffusivity $\alpha_p$ is essentially less than the electron thermal diffusivity $\alpha_e$. For this approximation the non-equilibrium temperature was being formed only in the electron subsystem interacted with the external laser pulse. Phonon temperature has remained to be equal to the temperature of the surrounding medium.

The objective of the present paper is to study, analytically the electron and phonon temperature distributions $T_{e,p}(x,t)$ and their time evolution taking into account the finite values for the phonon thermal conductivity and the electron thermal diffusivity. At the same time it is important to underline that we do not intend simply to calculate the small corrections to the temperatures $T_{e,p}$ obtained before. We will analyze the electron temperature behavior within a larger time interval compared to that of Ref. [5]. Besides, we are interesting in the influence of nonequilibrium electrons to the phonon temperature evolution. This problem is of interest because the electron and phonon temperatures can not be measured in the same way. For example, the phonon temperature can be measured by the calorimetric or photo-thermal methods but we can not apply these methods to the electron temperature measurements. That is why it is necessary to use other techniques, for instance thermoelectric measurements. The view of this we have to analyze each of these temperatures separately, no matter how small they might be. In addition, the temporal evolution of transient electron and phonon temperatures has different rates because of the different magnitudes of electron and phonon thermal parameters. Thus, if the electron temperature at some time is much greater than the phonon temperature, it means that this situation is valid only for a particular moment, and the temperature ratio may change later.

The separate studies of the electron and phonon temperature permit to obtain data about the thermal, optical, relaxation and other properties of the semiconductor quasiparticles. The comparison of the nonstationary phonon temperature in insulators and semiconductors allows us to understand the role of the electron subsystem in the formation of the nonstationary phonon temperature fields.

## II. Main Equations and General Results

Let us assume that a rectangular laser pulse with an arbitrary duration $\tau$ and intensity $I_0$, is inceding onto the left side $x=0$ of an isotropic, nondegenerate parallelepiped-shaped semiconductor with a unit section. The right side $x=l$ is kept at the constant ambient temperature $T_0$. The lateral sides of the sample are adiabatic insulated, so the problem is unidimensional. Furthermore, we assume for simplicity that the semiconductor is optically opaque. This means that the total laser energy irradiation is absorbed on the surface and is completely converted into heat. We consider the intensity $I_0$ to be so small that all the kinetic coefficients do not depend on the non-equilibrium temperatures.

With these assumptions the set of coupled   thermal diffusion equations for electrons and phonons become linear, and can be written in the form [7],

$$\begin{cases} \dfrac{\partial^2 T_e(x,t)}{\partial x^2} - k_e^2[T_e(x,t) - T_p(x,t)] = \dfrac{1}{\alpha_e} \dfrac{\partial T_e(x,t)}{\partial t} \\ \dfrac{\partial T_p(x,t)}{\partial x^2} + k_p^2[T_e(x,t) - T_p(x,t)] = \dfrac{1}{\alpha_p} \dfrac{\partial T_p(x,t)}{\partial t} \end{cases} \qquad (1)$$

where, $k_{e,p}^2 = \dfrac{n\nu_\varepsilon}{\kappa_{e,p}}$ [6], $n$ is the electron concentration, and $\nu_\varepsilon$ is the energy relaxation frequency under the electron-phonon interaction. Hereafter the temperature is given in the energy units.

The quantities $k_{e,p}^{-1}$ are the so-called electron and phonon energy relaxation lengths (cooling lengths). These lengths determine the distance $k^{-1} = \left( \dfrac{k_e k_p}{k_e + k_p} \right)^{-1}$ within which the difference between the electron and phonon temperatures disappears in steady states.

The initial and boundary conditions for Eq. (1) are,

$$T_{e,p}(x,t)\big|_{t \leq 0} = T_0 \,,$$

$$T_{e,p}(x,t)\big|_{x=l} = T_0 \,,$$

$$-\kappa_{e,p} \frac{\partial T_{e,p}(x,t)}{\partial x}\bigg|_{x=0} = Q_{e,p}^0 \qquad (0 \leq t \leq \tau)\,, \qquad\qquad (2)$$

$$\frac{\partial T_{e,p}(x,t)}{\partial x}\bigg|_{x=0} = 0 \qquad\qquad (\tau \leq t < \infty)\,,$$

$$T_{e,p}(x, t = \tau - 0) = T_{e,p}(x, t = \tau + 0)\,.$$

The values $Q_{e,p}^0$ represent the incoming thermal fluxes to the electron and phonon gases as a result of the complete optical-to-heat surface energy conversion.

They are the phenomenological parameters in the present work. In fact, only the free-

charge carriers absorb the laser pulse, i.e. the total energy of the pulse is transferred

into heat of the electron gas. Then, a part of this heat is transferred to the phonon

subsystem via the surface and bulk electron-phonon energy interaction, thus causing

the heating of the phonon gas. We describe the surface heating by the term $Q_p^0$

which appears in the boundary condition (2). It is clear that $Q_e^0 > Q_p^0$. For simplicity of the model, we also assume that the recombination occurring on the illumination surface is strong enough for all the nonequilibrium carriers. Thus, they recombine instantaneously, and the released energy is transferred into heat.

The general solution of Eq. (1) for a semiconductor with an arbitrary degree of degeneration can be found in the Appendix. Here, we present this solution for nondegenerate semiconductors ($\kappa_e << \kappa_p$, $\alpha_e >> \alpha_p$), considering finite values for $\kappa_p$ and $\alpha_e$.

The electron and phonon temperatures for the time interval $0 \leq t \leq \tau$ are represented by the following expressions:

$$T_e(x,t) = T_0 + \frac{Q_e^0 + Q_p^0}{\kappa_p}(l-x) + (\frac{Q_e^0}{\kappa_e} - \frac{Q_p^0}{\kappa_p})\frac{sh[k_e(l-x)]}{k_e ch(k_e l)}$$

$$- \frac{2Q_e^0 l}{\kappa_e}\sum_{n=0}^{\infty}\frac{\cos(\beta_n \frac{x}{l})}{k_e^2 l^2 + \beta_n^2}e^{-\frac{t}{\tau_{en}}} - 2k_e^2 l^3 \frac{Q_e^0 + Q_p^0}{\kappa_p}\sum_{n=0}^{\infty}\frac{\cos(\beta_n \frac{x}{l})}{\beta_n^2(k_e^2 l^2 + \beta_n^2)}e^{-\frac{t}{\tau_{pn}}}, \qquad (3a)$$

$$T_p(x,t) = T_0 + \frac{Q_e^0 + Q_p^0}{\kappa_p}(l-x) - \frac{\kappa_e}{\kappa_p}\left(\frac{Q_e^0}{\kappa_e} - \frac{Q_p^0}{\kappa_p}\right)\frac{sh[k_e(l-x)]}{k_e ch(k_e l)}$$

$$- \frac{2Q_e^0 k_e^2 l^3}{\kappa_p}\sum_{n=0}^{\infty}\frac{\cos(\beta_n \frac{x}{l})}{\beta_n^2(k_e^2 l^2 + \beta_n^2)}e^{-\frac{t}{\tau_{pn}}} - \frac{2Q_p^0 l}{\kappa_p}\sum_{n=0}^{\infty}\frac{\cos(\beta_n \frac{x}{l})}{\beta_n^2}e^{-\frac{t}{\tau_{pn}}}. \qquad (3b)$$

Here, $\beta_n = \frac{\pi}{2}(2n+1)$, $n = 0,1,2...$, $\tau_{en} = \frac{l^2}{\alpha_e(k_e^2 l^2 + \beta_n^2)}$, $\tau_{pn} = \frac{l^2}{\alpha_p \beta_n^2}$.

After the pulse action is finished ($t > \tau$), these temperatures become

$$T_e(x,t) = T_0 + \frac{2Q_e^0 l}{\kappa_e}\sum_{n=0}^{\infty}\frac{\cos(\beta_n \frac{x}{l})}{k_e^2 l^2 + \beta_n^2}\left(e^{\frac{\tau}{\tau_{en}}} - 1\right)e^{-\frac{t}{\tau_{en}}}$$

$$+ \frac{2Q_e^0 k_e^2 l^3}{\kappa_p}\left(1 + \frac{Q_p^0}{Q_e^0}\right)\sum_{n=0}^{\infty}\frac{\cos(\beta_n \frac{x}{l})}{\beta_n^2(k_e^2 l^2 + \beta_n^2)}\left(e^{\frac{\tau}{\tau_{pn}}} - 1\right)e^{-\frac{t}{\tau_{pn}}}, \qquad (4a)$$

$$T_p(x,t) = T_0 + \frac{2Q_e^0 k_e^2 l^3}{\kappa_p} \sum_{n=0}^{\infty} \frac{\cos(\beta_n \frac{x}{l})}{\beta_n^2 (k_e^2 l^2 + \beta_n^2)} \left( e^{\frac{\tau}{\tau_{pn}}} - 1 \right) e^{-\frac{t}{\tau_{pn}}}$$

$$+ \frac{2Q_p^0 l}{\kappa_p} \sum_{n=0}^{\infty} \frac{\cos(\beta_n \frac{x}{l})}{\beta_n^2} \left( e^{\frac{\tau}{\tau_{pn}}} - 1 \right) e^{-\frac{t}{\tau_{pn}}} . \tag{4b}$$

It follows from the results obtained that for the nondegenerate semiconductors the transient thermal processes in a phonon subsystem are described by the characteristic time

$$\tau_{p0} = \frac{4l^2}{\pi^2 \alpha_p} . \tag{5}$$

This time depends on the phonon thermal diffusivity and the sample length; no electron parameters are contained in it. The characteristic time in (5) coincides with the characteristic relaxation time of heat processes in insulators (see Ref. [4]).

Unlike the phonons, the relaxation process in the electron gas is described by two characteristic times. The first of them is

$$\tau_{e0} = \frac{l^2}{\alpha_e (k_e^2 l^2 + \frac{\pi^2}{4})} , \tag{6}$$

and it is determined by the electron thermal diffusivity, the sample length and the correlation between the sample and the electron cooling lengths. This means that the nature of the electron energy relaxation is different for thin layers ($k_e l \ll 1$) and for thick samples ($k_e l \gg 1$).

The second electron relaxation time is the above mentioned time $\tau_{p0}$.

Since $\alpha_e >> \alpha_p$, it is clear that $\tau_{e0} << \tau_{p0}$. The presence of two characteristic times leads to the following behavior of the electron transient temperature: it is much higher at the first stage of the relaxation process ($\tau \leq t << \tau_{e0}$) than the phonon temperature, and at the second stage ($t > \tau_{e0}$) the electron temperature $T_e$ is comparable with $T_p$ (and even less in some cases). From Eqs. (4) it follows that at the moment of "switching off" the pulse, the temperatures $T_{e,p}(x, t = \tau)$ are determined by the thermal conductivities $\kappa_{e,p}$, the boundary thermal fluxes $Q_{e,p}^0$, the correlation between the sample length and the electron cooling length $l_\varepsilon = k_e^{-1}$, and the correlation between the pulse duration and the characteristic times $\tau_{e0}$ and $\tau_{p0}$.

The electron thermal conductivity $\alpha_e$, the correlation between the sample and the cooling lengths, as well as the correlation between the pulse duration and the time $\tau_{e0}$, determine the behavior of the electron temperature relaxation at the interval $[\tau, \tau_{e0}]$. This process was analyzed in detail in Ref. [5]. After the expiration of this time, the rate of the electron temperature damping becomes abruptly slower and the time $\tau_{p0}$ becomes the characteristic time of the relaxation process. Now the phonon thermal diffusivity and the sample length determine the behavior of the electron relaxation temperature.

As for the phonon temperature, its relaxation nature is determined only by the time $\tau_{p0}$, which depends exclusively on phonon parameters. The amplitude of the phonon temperature is determined by the phonon and electron parameters.

It should be reminded that the phonon temperatures (3b, 4b) consist of two parts. One of them depends on the phonon parameters only, and coincides with the nonequilibrium temperature in insulator [4]. The second one depends on both the electron and phonon parameters and describes the influence of the electron subsystem on the nonstationary phonon temperature field.

## III. The Physical Interpretation of Phenomena

The exact analytical solutions obtained for the transient electron and phonon temperatures (3, 4) are very complex for understanding of their physical meaning. Therefore, let us consider the simplified expressions for the temperatures $T_{e,p}(x,t)$ which contain only zero harmonics. The main goal of this simplification is the qualitative understanding of the physical phenomenon (the account of the next harmonics and their numerical calculations will be done later on).

Further we will pay attention to the relaxation interval only ($t \geq \tau$) because it is the most interesting temporal interval from the physical point of view,.

For this interval, the approximate transient temperatures can be presented as follows

$$T_e(x,t) = T_0 + \frac{2Q_e^0 l}{\kappa_e} \frac{\cos(\frac{\pi x}{2l})}{k_e^2 l^2 + \frac{\pi^2}{4}} \left[ \left( e^{\frac{\tau}{\tau_{e0}}} - 1 \right) e^{-\frac{t}{\tau_{e0}}} + \frac{4(k_p l)^2}{\pi^2} (1 + \frac{Q_p^0}{Q_e^0}) \left( e^{\frac{\tau}{\tau_{p0}}} - 1 \right) e^{-\frac{t}{\tau_{p0}}} \right] \quad (7a)$$

$$T_p(x,t) = T_0 + \frac{8Q_p^0 l}{\pi^2 \kappa_p} \cos(\frac{\pi x}{2l}) \left( 1 + \frac{Q_e^0}{Q_p^0} \frac{k_e^2 l^2}{k_e^2 l^2 + \frac{\pi^2}{4}} \right) \left( e^{\frac{\tau}{\tau_{p0}}} - 1 \right) e^{-\frac{t}{\tau_{p0}}} \quad (7b)$$

The macroscopic time $\tau_{p0}$ determines the relaxation time of nonequilibrium process in the whole phonon subsystem. The mechanism of this relaxation is the thermal diffusion in the phonon gas. It is clear from Eq. (7b) that electrons have more influence on phonons if the sample length is $l \sim k_e^{-1}$, and if $Q_e^0 >> Q_p^0$.

The nature of the energy relaxation in the electron subsystem is much more complicated. As it can be observed from Eq. (6), in general at the high-speed relaxation interval ($\tau \leq t \approx \tau_{e0}$) electrons transfer their energies to the external reservoir in two different ways. One of the relaxation mechanisms is the internal thermal diffusion in the electron gas with the characteristic time $\tau_e = \dfrac{4l^2}{\pi^2 \alpha_e}$; in this case the situation is similar to phonons. The second one is the energy interaction with phonons with characteristic time $\tau_\varepsilon = \dfrac{l_\varepsilon^2}{\alpha_e}$. It should be mentioned that this last time coincides with the electron energy relaxation time, which is well known in the theory of hot electrons [8,9]. The phonon subsystem is the internal quasistatic heat reservoir for electrons in the last case.

In order to take into account this complicated relaxation mechanism we write down the generalized effective energy relaxation length

$$l^* = \sqrt{\frac{l^2 l_\varepsilon^2}{l^2 + l_\varepsilon^2}}, \tag{8}$$

which determines the thermal diffusion length of the energy interaction with both the internal and external heat reservoirs.

This length is equal to the sample length $l$ for the limit case of thin sample ($k_e l << 1$), and the only mechanism of the electron energy relaxation is the thermal diffusion. In the opposite limit case of the long samples $l^* = l_\varepsilon$ for thick samples ($k_e l \gg 1$). From the physical point of view this means that the electron thermal diffusion becomes noneffective in thick samples, and the electrons relax energy into the phonon subsystem through the electron-phonon interaction. The characteristic length of this process is the electron cooling length. Let us consider these situations in detail.

***Thin Layers*** *($k_e l \ll 1$).*

The characteristic relaxation time in the electron subsystem, as it follows from (6), is

$$\tau_{e0} = \tau_{ed} = \frac{4l^2}{\pi^2 \alpha_e} \ , \tag{9}$$

i.e. the energy relaxation takes place only by means of the electron thermal diffusion.

Therefore, if the sample length is small enough compared to the electron cooling length the correlation $k_p l \ll 1$ is satisfied automatically because $\kappa_p \gg \kappa_e$. This means that in thin layers both the electron and phonon cooling lengths are much greater than the sample length. Due to this fact the electron and phonon temperatures can be represented by the expressions

$$T_e(x,t) = T_0 + \frac{8 Q^0{}_e l}{\pi^2 \kappa_e} \cos(\frac{\pi x}{2l}) \left[ \left( e^{\frac{\tau}{\tau_{ed}}} - 1 \right) e^{-\frac{t}{\tau_{ed}}} + \frac{4(k_p l)^2}{\pi^2} \ (1 + \frac{Q^0{}_p}{Q^0{}_e}) \left( e^{\frac{\tau}{\tau_{p0}}} - 1 \right) e^{-\frac{t}{\tau_{p0}}} \right] \tag{10a}$$

$$T_p(x,t) = T_0 + \frac{8 Q^0{}_p l}{\pi^2 \kappa_p} \cos(\frac{\pi x}{2l}) \left( e^{\frac{\tau}{\tau_{p0}}} - 1 \right) e^{-\frac{t}{\tau_{p0}}} \tag{10b}$$

Since $\dfrac{\tau_{e0}}{\tau_{p0}} = \dfrac{\alpha_p}{\alpha_e} << 1$, then during the time $\tau_{e0}$ the second term in the square brackets of Eq. (10a) becomes negligibly small compared to the first one after the "switching off" of the pulse. Electrons rapidly relax their energy during this time into the external reservoir by means of the electron thermal diffusion (see the sketch in Fig. 1). This process should lead to the equality of the electron temperature to the equilibrium temperature $T_0$ of the external thermostat (dotted line in Fig.1). Practically the electron-phonon energy interaction becomes essential if $t > \tau_{ed}$. In this case the phonon subsystem acts as a bulk heater for the electron subsystem because after $\tau_{ed}$ the phonon temperature is higher than that of the electron temperature. The energy relaxation mechanism remains the same, electrons continue to relax non-equilibrium energy into the surrounding medium by means of the electron thermal diffusion.

It should be noted that in this case the electrons do not have any influence on the phonon thermal process because of the negligibly small role of the electron-phonon energy interaction. At this point, the heat transient process in the phonon gas is similar to the transient process in insulators [4].

### Thick Samples ($k_e l >> 1$).

The nature of the electron temperature relaxation is changed drastically when the relation $k_e l >> 1$ between the sample length and the cooling length takes place. Electrons transfer the excess of the nonequilibrium energy to phonons by means of only the electron-phonon energy interaction. This interaction occurs at the distance $\dfrac{1}{k_e} << l$ from the illuminated surface $x = 0$. Electrons are in the energy equilibrium state in the remaining sample volume. The relaxation time of this process coincides with the energy relaxation time at the electron-phonon interaction $\tau_{e0} = \tau_\varepsilon = \dfrac{l_\varepsilon^2}{\alpha_e}$.

The energy relaxation mechanism for phonons remains the same as before with the characteristic time $\tau_{p0} = \dfrac{4l^2}{\pi^2 \alpha_p} \gg \tau_\varepsilon$.

The temperature distributions for electrons and phonons for $t > \tau$ are the following,

$$T_e(x,t) = T_0 + \frac{2Q_e^0 l}{\kappa_e (k_e l)^2} \cos\left(\frac{\pi x}{2l}\right)$$

$$\left[ \left( e^{\frac{\tau}{\tau_\varepsilon}} - 1 \right) e^{-\frac{t}{\tau_\varepsilon}} + \frac{4(k_p l)^2}{\pi^2} \left( 1 + \frac{Q_p^0}{Q_e^0} \right) \left( e^{\frac{\tau}{\tau_{p0}}} - 1 \right) e^{-\frac{t}{\tau_{p0}}} \right], \tag{14a}$$

$$T_p(x,t) = T_0 + \frac{8(Q_p^0 + Q_e^0)l}{\pi^2 \kappa_p} \cos\left(\frac{\pi x}{2l}\right) \left( e^{\frac{\tau}{\tau_{p0}}} - 1 \right) e^{-\frac{t}{\tau_{p0}}}. \tag{14b}$$

It can be seen from Eqs. (14b) and (10b) that the phonon temperature is determined not only by the influence of the incoming phonon flux $Q_p^0$ but also by the electron flux $Q_e^0$ in thick samples. For this type of samples the other thermal electron parameters do not affect the temperature $T_p(x,t)$.

The electron temperature in thin layers is essentially less [by $(k_e l)^2$ times] than in thick samples.

In thin-film semiconductors the inequality $k_e l \ll 1$ automatically implies the inequality $k_p l \ll 1$. The situation is different for thick specimens. Now the phonon cooling length and the sample length may be in arbitrary correlations, from $k_p l << 1$ to $k_p l >> 1$.

This means that there are possible situations when the second term in the square brackets of Eq.(14a) is essentially less than the first one or, vice versa, is essentially greater than the first term.

If $(k_p l)^2 << \left( e^{\frac{\tau}{\tau_\varepsilon}} - 1 \right) \left( e^{\frac{\tau}{\tau_p}} - 1 \right)^{-1}$, then the electron temperature peak value is essentially greater than the phonon value $T_p(x, t = \tau)$ [see a sketch in Fig. 2]. In this case the phonon gas is also a quasistatic internal heat reservoir for electrons but with lower temperature than the electron gas, thus, being now an internal quasistatic cooler.

If the inequality $(k_p l)^2 >> \left( e^{\frac{\tau}{\tau_\varepsilon}} - 1 \right) \left( e^{\frac{\tau}{\tau_{p0}}} - 1 \right)^{-1}$ holds, then the electron and the phonon temperatures are equal and are determined by Eq. (14b).

This situation corresponds either to an extremely strong electron-phonon energy interaction ($l_\varepsilon \to 0$) or to the infinitely long sample ($l \to \infty$). In the latter case, the electron temperature differs from the phonon one within a very narrow layer adjoined to the surface x=0.

## IV. The Discussion of Numeric Calculations.

Some numerical results obtained within the framework of chosen approach will regard to Eq. (4) are given in Fig. 3-16. In order to illustrate the most important and universal results, we present the renormalized temperatures $\theta_{e,p}(\eta,\zeta) = \dfrac{T_{e,p}(x,t) - T_0}{2Q^0{}_e l}\kappa_e$, where $\eta = \dfrac{x}{l}$, $\xi = \dfrac{t}{\tau}$. In all the calculations we have taken into account that $\dfrac{\alpha_e}{\alpha_p} = 10^4$, $\dfrac{\kappa_e}{\kappa_p} = 10^{-3}$ [5], and $Q_e{}^0 = Q^0{}_p$.

The last correlation has been chosen for simplicity. The numerical calculations show that the deviation from it does not give any new physical results.

We can see from Eqs. (3,4) that the absolute values of temperatures essentially depend on the pulse duration. The pulse was defined in the problem with one type of the heat carrier as "long" when its duration $\tau$ is essentially greater than the single characteristic time $\tau_0$ [4]. In the opposite case $\tau \ll \tau_0$, the pulse was defined as "short". The temperature distributions are qualitatively different for both of these cases.

The situation becomes more diverse in the problem with two types of carriers of heat. The presence of two characteristic times ($\tau_{e0}, \tau_{p0}$) which are essentially different by scale ($\tau_{e0} \ll \tau_{p0}$) leads to the situation when for example the pulse can be long for the electron subsystem, and short for the phonon subsystem.

In general the following situations are possible:

1. The pulse is long with respect to both subsystems: $\tau \gg \tau_{p0} \gg \tau_{ed}$. In this case there are nonequilibrium quasistatic states for both electrons and phonons.

2. The pulse is long with respect to the electron subsystem and is short with respect to the phonon one: $\tau_{p0} \gg \tau \gg \tau_{e0}$. The quasistationary state can be achieved only in the electron gas.

3. The pulse is short with respect to both subsystems: $\tau_{p0} \gg \tau_{e0} \gg \tau$. None of the quasiparticle subsystems reaches quasistationary states under these conditions.

From the point of view of the study of dynamic heat processes, only the pulses with intermediate duration ($\tau_{e0} \ll \tau \ll \tau_{p0}$) and short pulses ($\tau \ll \tau_{e0} \ll \tau_{p0}$) are of physical interest. In fact, the case of long pulses corresponds to the static thermal fields.

The analysis will be carried out separately for thin and thick samples in analogy with the previous section. The samples will be characterized by the numerical data $k_e l = 0,1$ and $k_e l = 10$, respectively.

***Thin Layers.***

The temperature distributions $\theta_{e,p}(\eta, \zeta)$ with intermediate duration are shown in Figs. 3–6. It has to be reminded that the electron temperature is quasistatic, as it must be for a long pulse. Unlike the electron, after the "switching off" of the external perturbation the phonon relaxation process has a nonmonotonic-in-time nature. This is displayed through the appearance of the local heating of lattice in some region of the sample [see Figs. 5,6]. The physical reasons for different behavior of the two temperatures $\theta_e(\eta, \zeta)$ and $\theta_p(\eta, \zeta)$ is the following. Under a fixed duration of the perturbation pulse $\tau$ ($\tau \gg \tau_{ed}$) electrons can be heated within the length $l_{e\tau} = \sqrt{\alpha_e \tau}$ which is much greater than the sample length $l$ ($l = \sqrt{\alpha_e \tau_{ed}}$). Therefore, the electron gas is heated in the whole volume of the sample.

The situation is opposite for the phonon gas: the length of the phonon heating is $l_{p\tau} = \sqrt{\alpha_p \tau}$, and it is essentially less than the sample length $l$ $(l = \sqrt{\alpha_p \tau_{p0}})$. That is why only some part of the phonon gas close to the surface x=0 during the heat pulse is in the nonequilibrium state with the temperature $T_p > T_0$. After the pulse action, this region begins to cool, and the non-heated region is heated because of the thermal flux flowing into it. Some time later, the whole phonon subsystem reaches the equilibrium state.

In the case of short pulse with respect to both quasiparticular subsystems $(\tau \ll \tau_{ed} \ll \tau_{p0})$, the nature of the electron temperature changes qualitatively and becomes nonmonotonic in time according to the aforementioned arguments (Fig. 7,8). In principle, the nature of the phonon temperature does not change under this time hierarchy.

***Thick Samples.***

As it has already been noted, the electron-phonon energy interaction is the dominating energy relaxation mechanism of electrons in thick semiconductors, and this interaction is effective only within the cooling length $l_\varepsilon$. It means that only the electrons located near the surface x=0 within this length are able to reach a nonequilibrium state due to the laser absorption. Therefore, the criterion of the total warming-up of these electrons is

$$\sqrt{\alpha_e \tau} > l_\varepsilon . \tag{15}$$

This result is demonstrated in Figs.9.10.

If the inverse correlation $\sqrt{\alpha_e \tau} < l_\varepsilon$ takes place, the regions of local heating appear in the electron subsystem (Figs. 11, 12).

The situation in the phonon subsystem for a thick sample does not change qualitatively in comparison with short samples [see, e.g. Figs. 13, 14. The pulse has an intermediate duration].

The transient electron and phonon temperature distributions are shown in Figs. 15, 16 in the case of very long samples or very strong electron-phonon energy interaction ($k_e l = 10^3$). As a confirmation of a qualitative discussion given in the previous section, these electron and phonon temperatures coincide.

In conclusion of this section let us note that the absolute values of the temperatures $\theta_{e,p}$ are changed substantially if both the sample length and the pulse duration change. In the general case these changes lead to the following conclusion: both the electron and phonon temperatures decrease when the sample length increases and if besides the pulse duration decreases. Therefore, the electron temperature at the moment of "switching off" of the pulse essentially exceeds the phonon temperature at the same moment of time.

**V. Conclusions**

The electron and phonon transient temperature distributions in nondegenerate semiconductors have been studied within the framework of a two-temperature model. Finite values of the electron-phonon energy interaction and finite values of the electron thermal diffusivity and the phonon thermal conductivity have been taken into account.

It is shown that the phonon thermal diffusion is the dominating mechanism of the energy relaxation in the phonon gas. The energy relaxation mechanism is different for the electron subsystem and depends on the correlation between the sample length and the electron cooling length. The electron thermal diffusion is the main mechanism for thin layers, while for thick samples it is the electron-phonon energy interaction.

The electron temperature is determined by two characteristic times at the relaxation interval for both thin and thick specimens. One of these times includes the electron thermal diffusivity, and determines the initial, high-speed interval of the nonequilibrium temperature. The second time includes phonon thermal diffusivity and determines the smooth relaxation process of the temperature $T_e$. The possibility of obtaining the electron and phonon parameters for single experiment arises from the analysis of responses (for example, thermoelectric response) to these two temperature relaxation intervals.

The difference between the values of relaxation times leads to a different correlation between these times and the duration of perturbation pulse. Therefore, in some cases the situations are possible when the same pulse which is of short duration with respect to one quasiparticular subsystem, turns out to be long with respect to the other subsystem. As a result, qualitatively different temperature distributions are produced in both gases.

We have also analyzed the influence of the electron subsystem on the phonon relaxation process. This influence is more essential in the case of the samples which have the lengths $l$ comparable with the electron cooling length ($k_e l \approx 1$).

**Appendix**

To obtain the general solution of  Eq.(1) with the initial and boundary conditions (2), it is convenient to divide the problem  into  two time intervals: the time of pulse action ($0 \leq t \leq \tau$),  and the time after the "switching off" of the pulse ($t > \tau$).

Therefore, the solution for the first interval at the moment $t = \tau$ is the initial condition for the solution of the problem for the second time interval.

**Interval** $0 \le t \le \tau$. First of all, let us represent the general solution of Eq.(1) as the sum of the static and dynamic components of the electron and phonon nonequilibrium temperatures,

$$T_{e,p}(x,t) = T_{e,p}^s(x) + F_{e,p}(x,t). \tag{A1}$$

The static temperatures satisfy the equations

$$\begin{cases} \dfrac{d^2 T_e^s}{dx^2} - k_e^2 \left(T_e^s - T_p^s\right) = 0, \\ \dfrac{d^2 T_p^s}{dx^2} + k_p^2 \left(T_e^s - T_p^s\right) = 0 \end{cases}, \tag{A2}$$

with the following boundary conditions:

$$-\kappa_{e,p} \left. \dfrac{dT_{e,p}^s}{dx} \right|_{x=0} = Q_{e,p}^0, \tag{A3}$$

$$T_{e,p}^s \big|_{x=l} = T_0. \tag{A4}$$

Eqs.(A2) are solved trivially, and the solution being the following functions,

$$T_{e,p}^s(x,t) = T_0 + \frac{1}{k^2}\left(\frac{Q_e^0}{\kappa_e}k_p^2 + \frac{Q_p^0}{\kappa_p}k_e^2\right)(l-x) \pm \frac{k_{e,p}^2}{k^2}\left(\frac{Q_e^0}{\kappa_e} - \frac{Q_p^0}{\kappa_p}\right)\frac{sh[k(l-x)]}{k\,ch(kl)} . \tag{A5}$$

The dynamical components of the temperatures $F_{e,p}(x,t)$ satisfy Eq.(1) when the functions $T_{e,p}(x,t)$ are substituted by the functions $F_{e,p}(x,t)$. The boundary and initial conditions change:

$$\left. \frac{\partial F_{e,p}(x,t)}{\partial x} \right|_{x=0} = 0, \tag{A6}$$

$$F_{e,p}\big|_{x=l} = 0, \tag{A7}$$

$$F_{e,p}(x,t)\big|_{t\leq 0} = T_0 - T_{e,p}^s(x).$$  (A8)

Let us present the dynamic temperature in terms of a Fourier series expansion which can be taken in the form

$$F_{e,p}(x,t) = \sum_n A_{e,p}^n(t)\cos\left(\beta_n\frac{x}{l}\right),$$  (A9)

due to (A6). Here $A_{e,p}^n(t)$ are the time dependent unknown coefficients, $\beta_n$ are the eigenvalues. The latter are easily obtained from condition (A7), and are equal to

$$\beta_n = \frac{\pi}{2}(2n+1) , \quad n = 0,1,..$$  (A10)

The coefficients $A_{e,p}^n(t)$ satisfy the following equations:

$$\begin{cases} \dfrac{dA_e^n}{dt} = a_1 A_e^n + a_2 A_p^n, \\ \dfrac{dA_p^n}{dt} = a_3 A_e^n + a_4 A_p^n \end{cases},$$  (A11)

where $a_1 = -\dfrac{\alpha_e}{l^2}\left(k_e^2 l^2 + \beta_n^2\right)$, $a_2 = \alpha_e k_e^2$, $a_3 = \alpha_p k_p^2$, $a_4 = -\dfrac{\alpha_p}{l^2}\left(k_p^2 l^2 + \beta_n^2\right)$.

We will look for the unknown values $A_{e,p}^n(t)$ in the following form:

$$A_{e,p}^n = b_{e,p}^n \exp(\lambda_n t) ,$$  (A12)

where $b_{e,p}^n$, $\lambda_n$ are the new unknown coefficients.

After substituting Eq.(A12) into Eq.(A11) we obtain a homogeneous algebraic set of equations for the unknowns $b_{e,p}^n$. To find the nontrivial solutions, it is necessary to equate the determinant of these combined equations to zero. Then, from this condition, we find two different values of $\lambda_n$:

$$\lambda_{1,2}^n = -\frac{1}{2}\left[\alpha_e k_e^2 + \alpha_p k_p^2 + \frac{\alpha_e + \alpha_p}{l^2}\beta_n^2 \mp \sqrt{\left[\alpha_e k_e^2 - \alpha_p k_p^2 + \frac{\alpha_e - \alpha_p}{l^2}\beta_n^2\right]^2 + 4\alpha_e\alpha_p k_e^2 k_p^2}\right]$$

(A13)

In this case we can rewrite Eq.(A12) as follows:

$$\begin{cases} A_e^n = b_{e1}^n \exp(\lambda_{1n}t) + b_{e2}^n \exp(\lambda_{2n}t), \\ A_p^n = b_{p1}^n \exp(\lambda_{1n}t) + b_{p2}^n \exp(\lambda_{2n}t). \end{cases}$$

(A14)

To obtain the four unknown coefficients $b_{ei,pi}^n$ (i=1,2), it is necessary to use the above mentioned algebraic combined equations and the two boundary conditions (A8). The first of them gives the possibility to express two of the arbitrary coefficients in terms of the others. The two remaining independent coefficients can be found from Eq.(A8). To do this, it is necessary to expand Eq.(A5) in terms of a Fourier series, and to substitute this expansion together with Eq.(A9) into Eq.(A8). Comparing the terms with the same $\cos(\beta_n \frac{x}{l})$ we obtain the new set of algebraic equations to determinate the last unknown values. As a result, the general solution of Eq.(1) takes the form:

$$T_e(x,t) = T_0 + \frac{1}{k^2}\left(\frac{Q_e^0}{\kappa_e}k_p^2 + \frac{Q_p^0}{\kappa_p}k_e^2\right)(l-x) +$$

$$+ \frac{k_e^2}{k^2}\left(\frac{Q_e^0}{\kappa_e} - \frac{Q_p^0}{\kappa_p}\right)\frac{\mathrm{sh}[k(l-x)]}{k\,\mathrm{ch}(kl)} + \frac{2k_e^2}{lk^2}\left(\frac{Q_e^0}{\kappa_e} - \frac{Q_p^0}{\kappa_p}\right)\times$$

$$\times \sum_{n=0}^{\infty}\cos\left(\beta_n\frac{x}{l}\right)\left[\frac{(k^2l^2\alpha_e + \beta_n^2\alpha_e + \lambda_{1n}l^2)e^{\lambda_{2n}t} - (k^2l^2\alpha_e + \beta_n^2\alpha_e + \lambda_{2n}l^2)e^{\lambda_{1n}t}}{(\lambda_{2n} - \lambda_{1n})(k^2l^2 + \beta_n^2)}\right] +$$

$$+ \frac{2}{lk^2}\left(\frac{Q_e^0}{\kappa_e}k_p^2 + \frac{Q_p^0}{\kappa_p}k_e^2\right)\times\sum_{n=0}^{\infty}\frac{\cos\left(\beta_n\frac{x}{l}\right)}{\beta_n^2}\left[\frac{(\beta_n^2\alpha_e + \lambda_{1n}l^2)e^{\lambda_{2n}t} - (\beta_n^2\alpha_e + \lambda_{2n}l^2)e^{\lambda_{1n}t}}{\lambda_{2n} - \lambda_{1n}}\right]$$

(A15)

$$T_p(x,t) = T_0 + \frac{1}{k^2}\left(\frac{Q_e^0}{\kappa_e}k_p^2 + \frac{Q_p^0}{\kappa_p}k_e^2\right)(l-x) -$$

$$-\frac{k_p^2}{k^2}\left(\frac{Q_e^0}{\kappa_e} - \frac{Q_p^0}{\kappa_p}\right)\frac{\text{sh}[k(l-x)]}{k\,\text{ch}(kl)} - \frac{2k_p^2}{lk^2}\left(\frac{Q_e^0}{\kappa_e} - \frac{Q_p^0}{\kappa_p}\right)\times$$

$$\times\sum_{n=0}^{\infty}\cos\left(\beta_n\frac{x}{l}\right)\left[\frac{\left(k^2l^2\alpha_p + \beta_n^2\alpha_p + \lambda_{1n}l^2\right)e^{\lambda_{2n}t} - \left(k^2l^2\alpha_p + \beta_n^2\alpha_p + \lambda_{2n}l^2\right)e^{\lambda_{1n}t}}{\left(\lambda_{2n} - \lambda_{1n}\right)\left(k^2l^2 + \beta_n^2\right)}\right] +$$

$$+\frac{2}{lk^2}\left(\frac{Q_e^0}{\kappa_e}k_p^2 + \frac{Q_p^0}{\kappa_p}k_e^2\right)\times\sum_{n=0}^{\infty}\frac{\cos\left(\beta_n\frac{x}{l}\right)}{\beta_n^2}\left[\frac{\left(\beta_n^2\alpha_p + \lambda_{1n}l^2\right)e^{\lambda_{2n}t} - \left(\beta_n^2\alpha_p + \lambda_{2n}l^2\right)e^{\lambda_{1n}t}}{\lambda_{2n} - \lambda_{1n}}\right]$$

$$(A16)$$

The solution for the interval $t \geq \tau$ can be determined in the same manner as it has been done for the interval $0 \leq t \leq \tau$. Eqs.(A15) and (A16) at the moment $t = \tau$ are the initial conditions for this problem.

Usually, in nondegenerate semiconductors, $\frac{\kappa_p}{\kappa_e} \approx 10^{-3}$ and $\frac{\alpha_e}{\alpha_p} \approx 10^4 \div 10^6$ [5]. The presence of small parameters $\frac{\kappa_e}{\kappa_p}$ and $\frac{\alpha_p}{\alpha_e}$ gives the possibility to simplify the expressions of $k^2$ and $\lambda_{in}$ (i=1,2): $k^2 \approx k_e^2$,

$$\lambda_{1n} \approx -\frac{\alpha_p}{l^2}, \quad \lambda_{2n} \approx -\frac{\alpha_e}{l^2}\left(k_e^2l^2 + \beta_n^2\right) \qquad (A17)$$

As a result, we obtain Eqs. (3) and (4).

### Acknowledgements

This work has been partially supported by the Consejo Nacional de Ciencia y Tecnologia (CONACYT), Mexico.

# REFERENCES

1. A. Othonos, J. Appl. Phys., **83,** 1789 (1998)

2. M. Sasaki, H. Negishi, and M. Jnoue, J. Appl. Phys., **59,** 796 (1986)

3.V. A. Kulbachinski, Z.M. Dashevski, M. Jnoue, M. Sasaki, H. Negishi, W.X. Gao, P. Lostak, J. Horak, and A. de Visser, Phys. Rev., **B52**, 10915 (1995)

4. Alvaro F .Carballo-Sanchez and Yu.G. Gurevich, G.N. Logvinov and Yu.V. Drogobitskiy, O.Yu. Titov, Physics of the Solid State, **41**, 544 (1999)

5. A.F. Carballo-Sanchez, G. Gonzalez de la Cruz, Yu.G. Gurevich and G.N. Logvinov, Phys.Rev. **B59**,10630 (1999)

6.Yu.G. Gurevich and O.L. Mashkevich, Phys.Rep.,**181**, 327 (1989)

7. G. Gonzalez de la Cruz and Yu.G. Gurevich, J.Appl.Phys., **80**, 1726 (1996)

8.Esther M.Conwell, High Field Transport in Semiconductors, Academic Press, New York and London, 1967.

9. F.G. Bass, Yu.G. Gurevich, Sov. Phys. Uspekhi, **14**, 113, (1971).

**FIGURE CAPTIONS**

1. Schematic behavior of electron and phonon temperatures in short samples.

2. Schematic behavior of electron and phonon temperatures in long samples.

3. The dependence of temperature on coordinate in short samples ($t \geq \tau$),

$\tau_{ed} \ll \tau \ll \tau_{p0}$

4. The dependence of temperature $\theta_e$ on time in short samples ($t \geq \tau$),

$\tau_{ed} \ll \tau \ll \tau_{p0}$

5. The dependence of temperature $\theta_p \cdot 10^{-5}$ on coordinate in the layers ($t \geq \tau$),

$\tau_{ed} \ll \tau \ll \tau_{p0}$

6. The dependence of temperature $\theta_p \cdot 10^{-5}$ on time in short samples ($t \geq \tau$),

$\tau_{ed} \ll \tau \ll \tau_{p0}$

7. The dependence of temperature $\theta_e \cdot 10^{-2}$ on coordinate in short samples ($t \geq \tau$);    $\tau \ll \tau_{ed} \ll \tau_p$

8. The dependence of temperature $\theta_e \cdot 10^{-2}$ on time in short samples ($t \geq \tau$);

$\tau \ll \tau_{ed} \ll \tau_p$

9. The dependence of temperature $\theta_e \cdot 10^{-2}$ on coordinate in long samples ($t \geq \tau$);    $\tau_\varepsilon \ll \tau \ll \tau_{p0}$

10 The dependence of temperature $\theta_e \cdot 10^{-2}$ on time in long samples ($t \geq \tau$);    $\tau_\varepsilon \ll \tau \ll \tau_{p0}$

11 The dependence of temperature $\theta_e \cdot 10^{-3}$ on coordinate in long samples ($t \geq \tau$);    $\tau \ll \tau_\varepsilon \ll \tau_p$

12. The dependence of temperature $\theta_e \cdot 10^{-3}$ on time in long samples ($t \geq \tau$);

$\tau \ll \tau_\varepsilon \ll \tau_p$

13. The dependence of temperature $\theta_p \cdot 10^{-4}$ on coordinate in long samples ($t \geq \tau$);      $\tau_\varepsilon \ll \tau \ll \tau_{p0}$

14. The dependence of temperature $\theta_p \cdot 10^{-4}$ on time in long samples ($t \geq \tau$);

$\tau_\varepsilon \ll \tau \ll \tau_{p0}$

15. The dependence of temperature $\theta_e \cdot 10^{-4}$ and $\theta_p \cdot 10^{-4}$ on coordinate in long samples under condition $k_e l = 10^3$ ($t \geq \tau$);      $\tau_\varepsilon \ll \tau \ll \tau_p$

16. The dependence of temperature $\theta_e \cdot 10^{-4}$ and $\theta_p \cdot 10^{-4}$ on time in long samples under condition $k_e l = 10^3$ ($t \geq \tau$);      $\tau_\varepsilon \ll \tau \ll \tau_p$

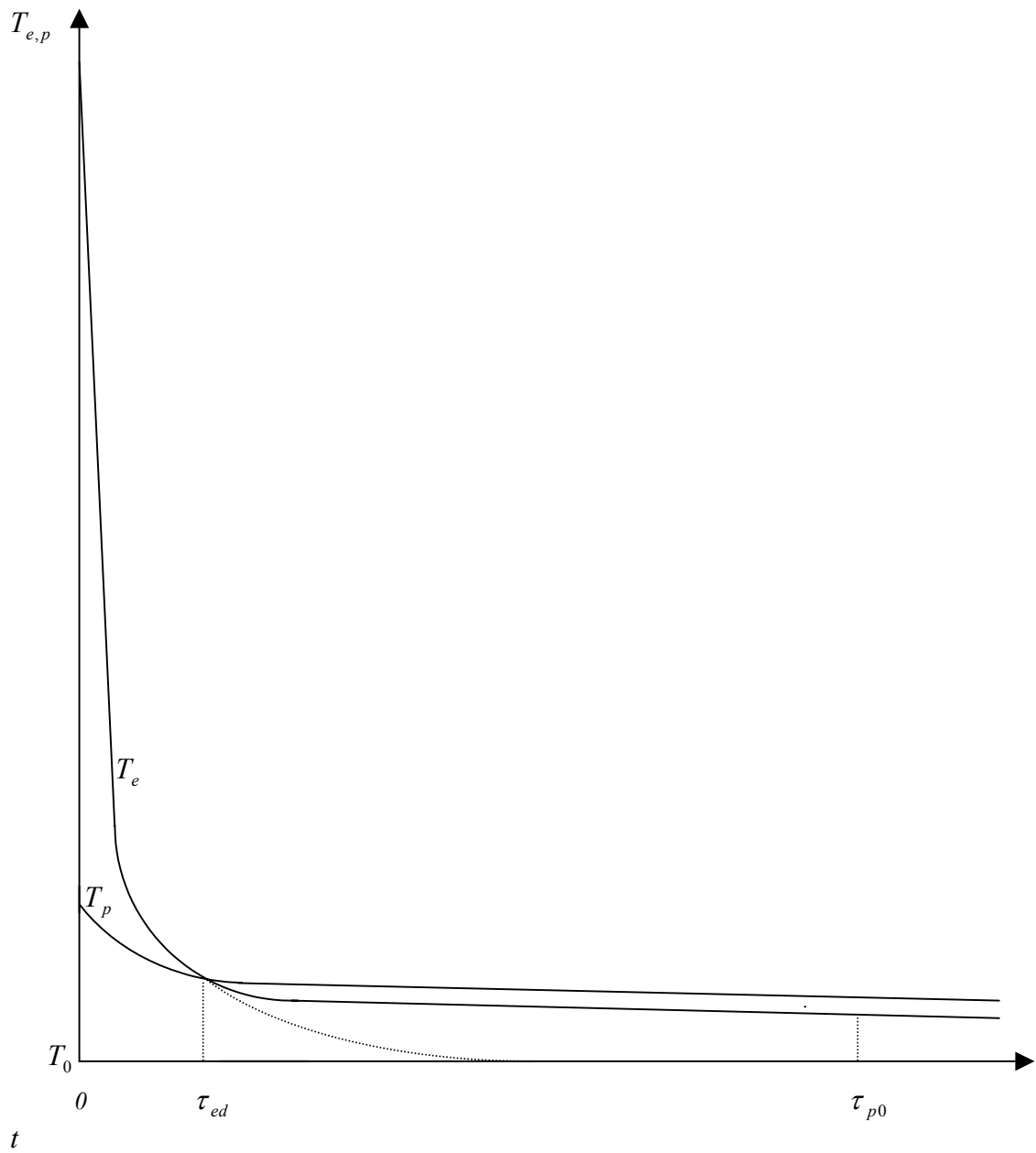

Fig. 1.  Yu. G. Gurevich, J. Appl. Phys.

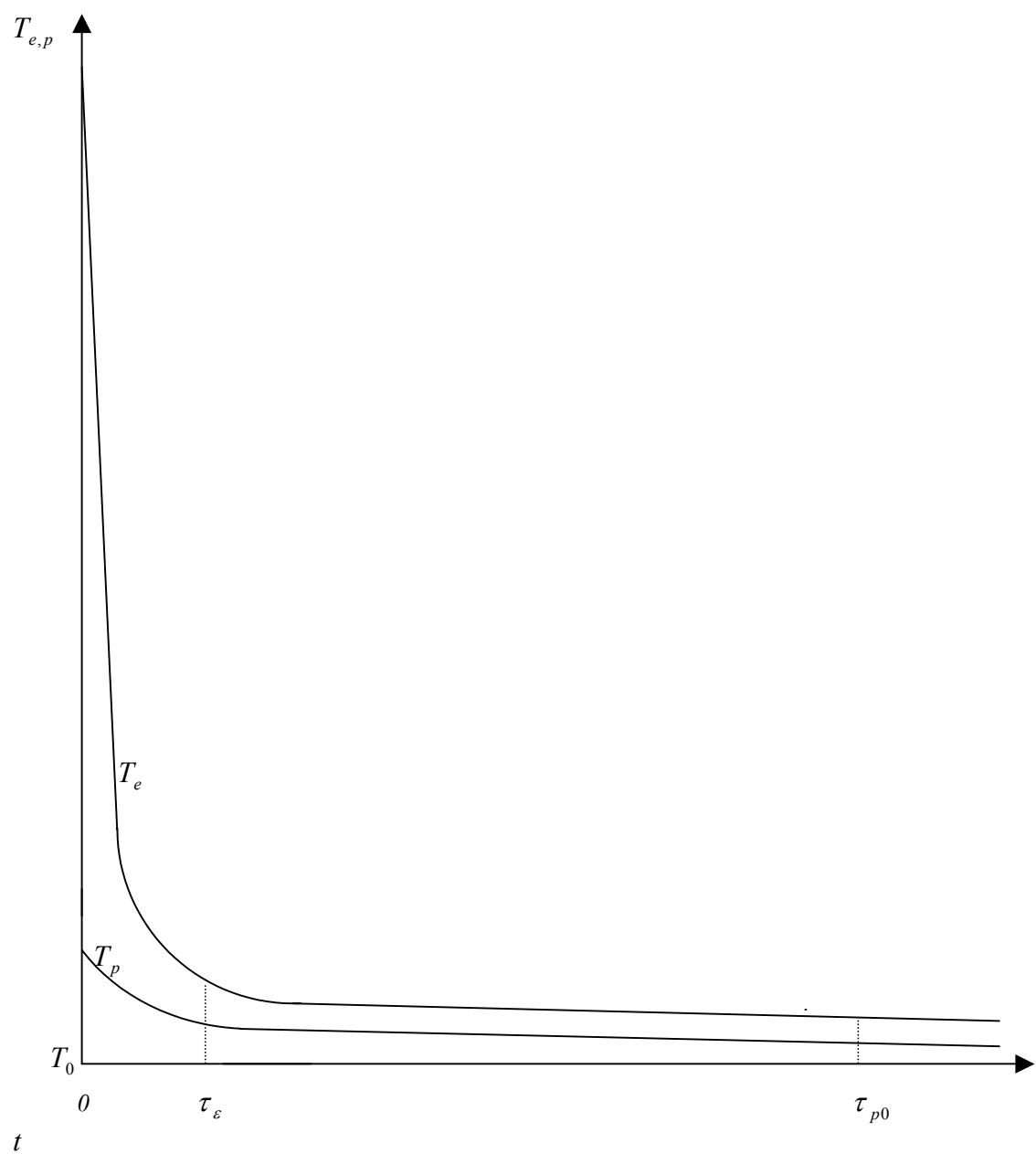

*Fig. 2. Yu. G. Gurevich, J. Appl. Phys.*

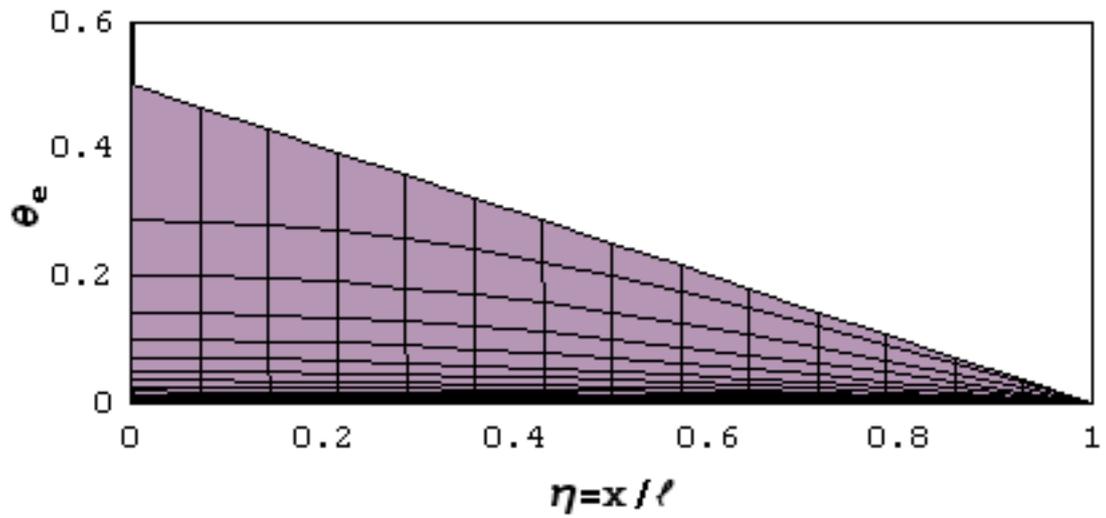

*Fig. 3. Yu. G. Gurevich, J. Appl. Phys.*

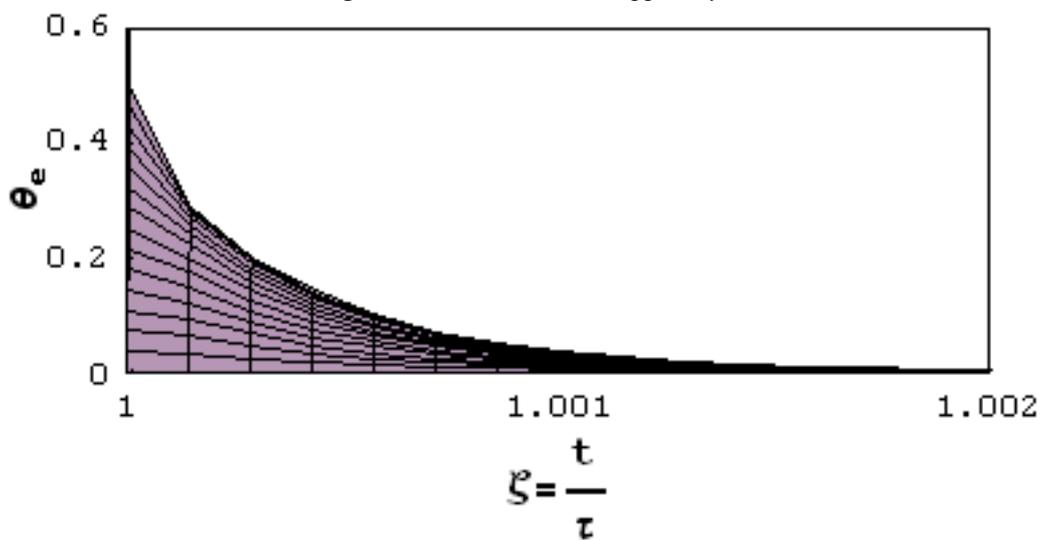

*Fig. 4. Yu. G. Gurevich, J. Appl. Phys.*

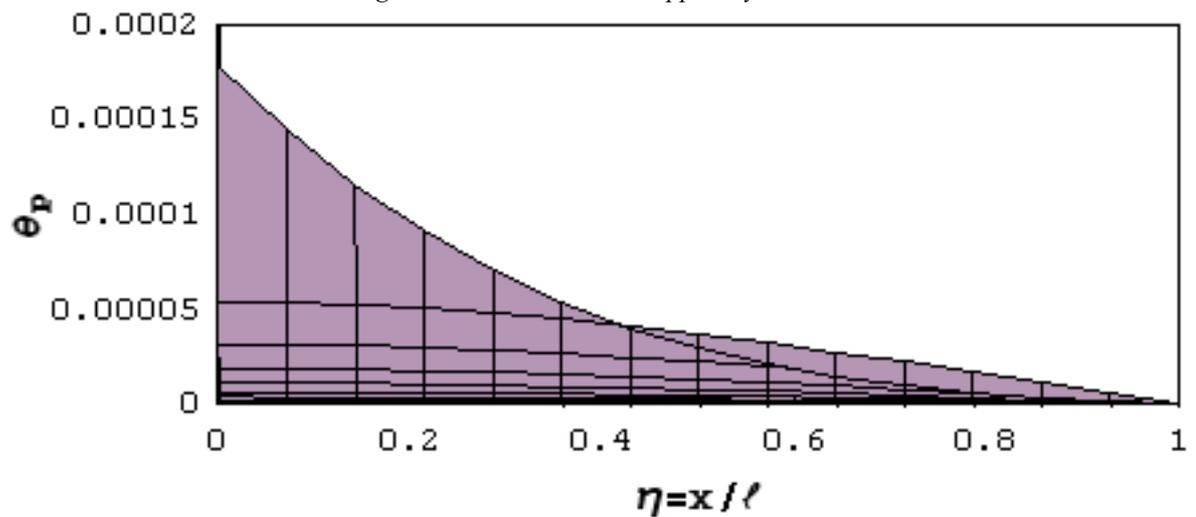

*Fig. 5. Yu. G. Gurevich, J. Appl. Phys.*

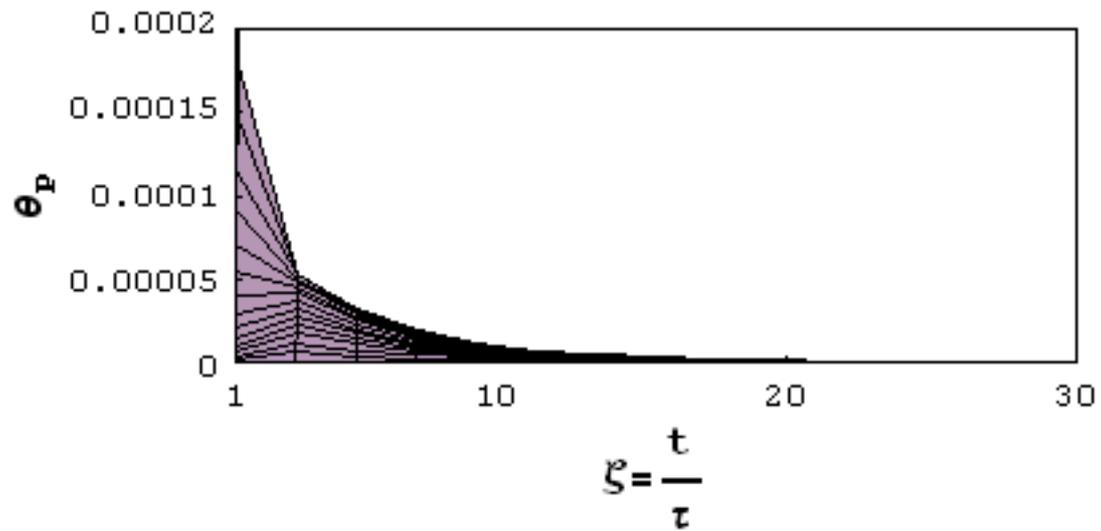

*Fig. 6. Yu. G. Gurevich, J. Appl. Phys.*

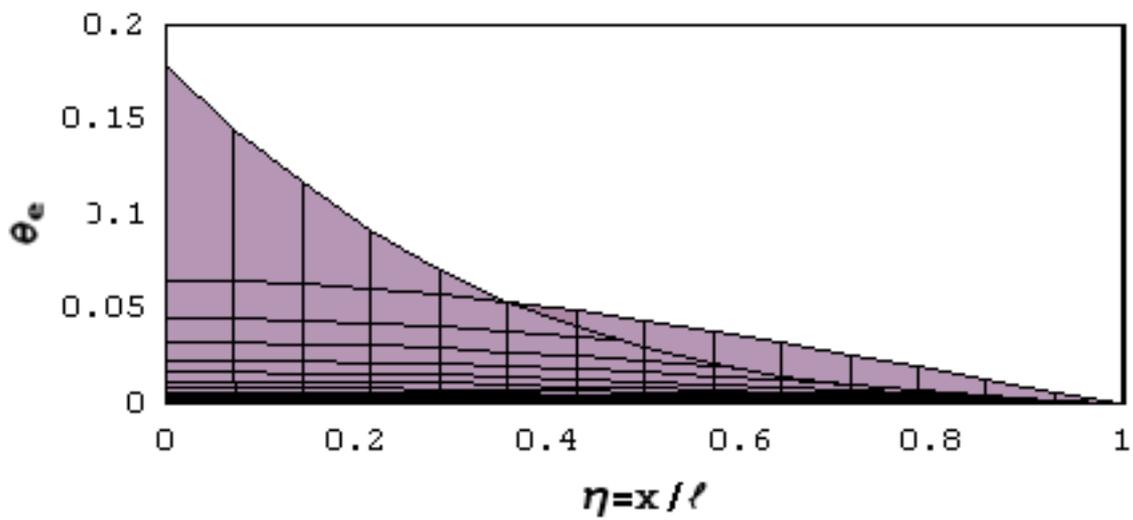

*Fig. 7. Yu. G. Gurevich, J. Appl. Phys*

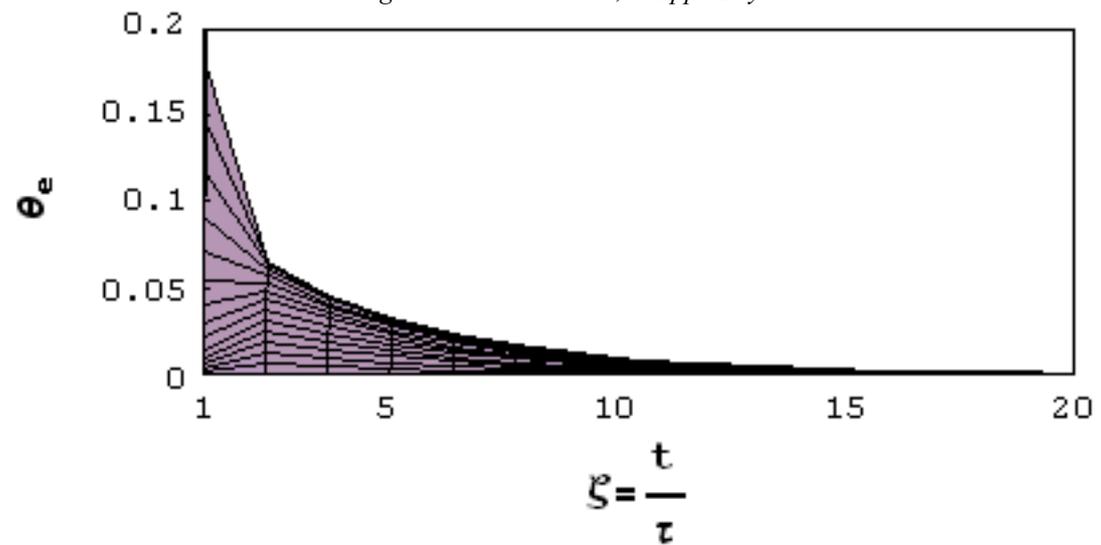

*Fig. 8. Yu. G. Gurevich, J. Appl. Phys.*

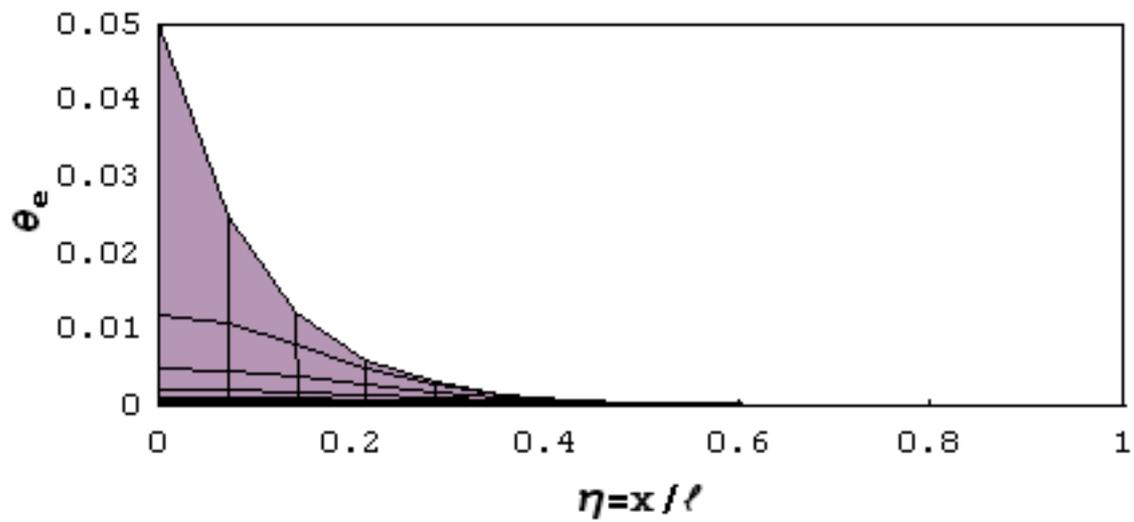

*Fig. 9.  Yu. G. Gurevich, J. Appl. Phys.*

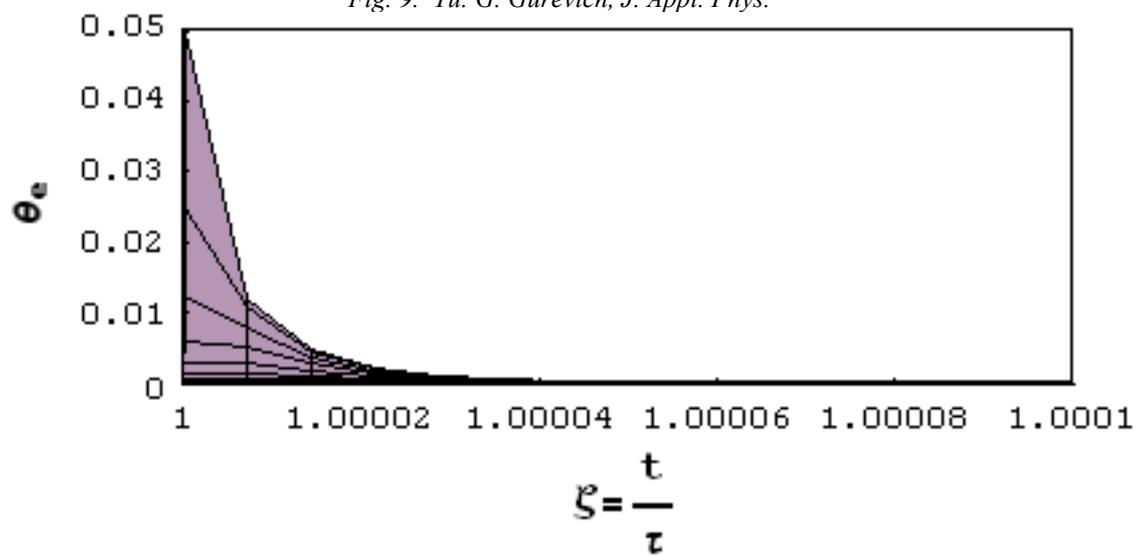

*Fig. 10.  Yu. G. Gurevich, J. Appl. Phys.*

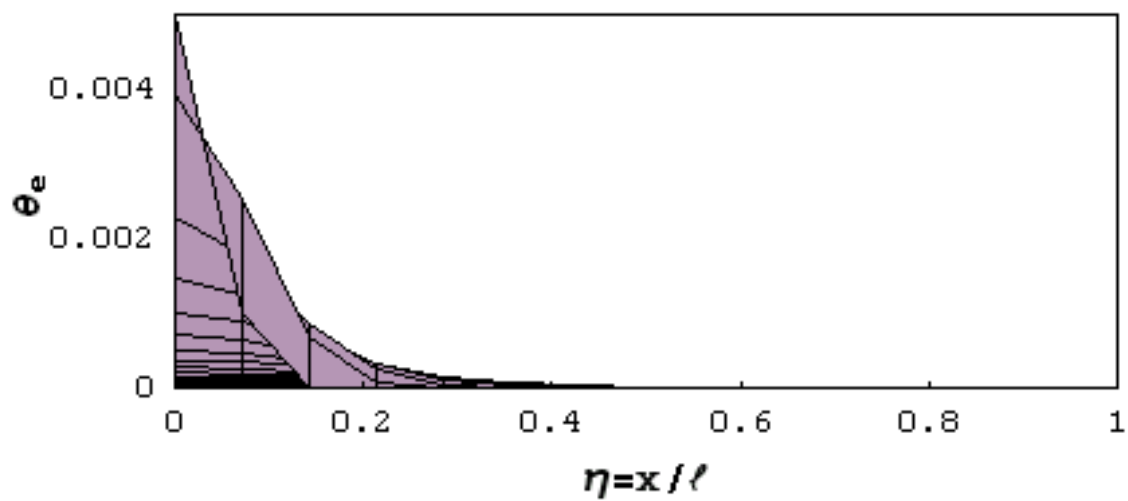

*Fig. 11.  Yu. G. Gurevich, J. Appl. Phys.*

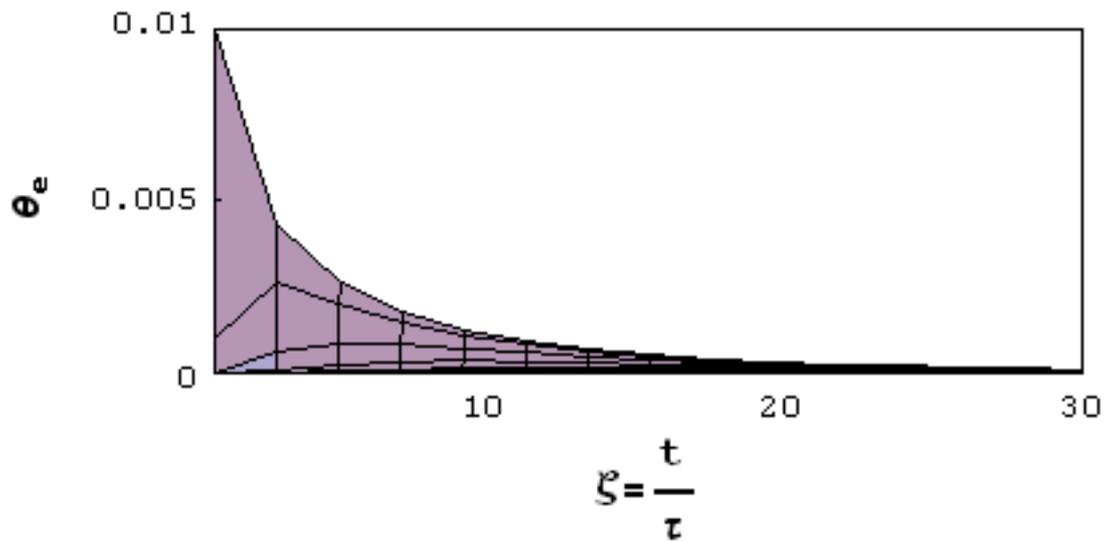

*Fig. 12. Yu. G. Gurevich, J. Appl. Phys.*

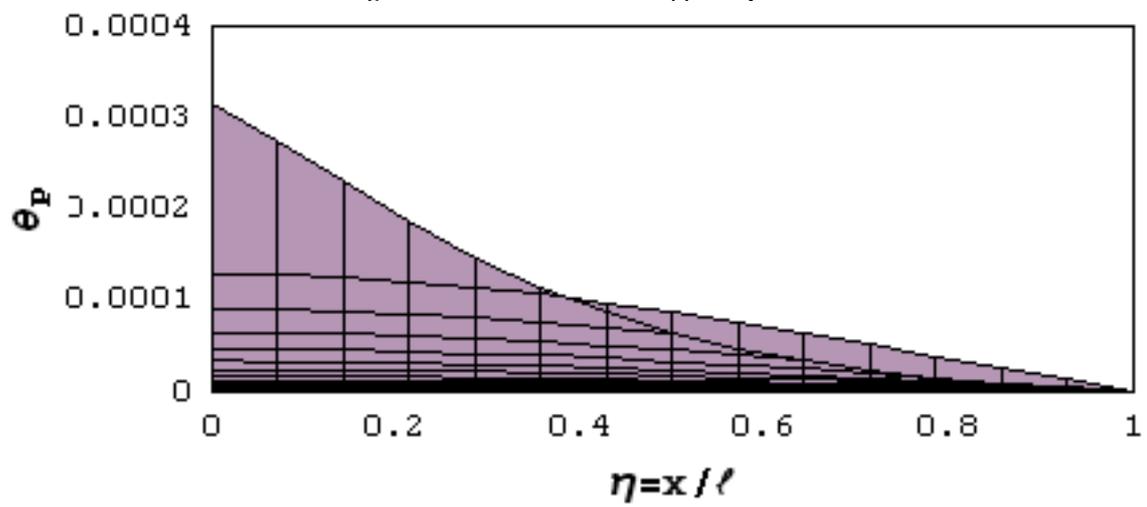

*Fig. 13. Yu. G. Gurevich, J. Appl. Phys.*

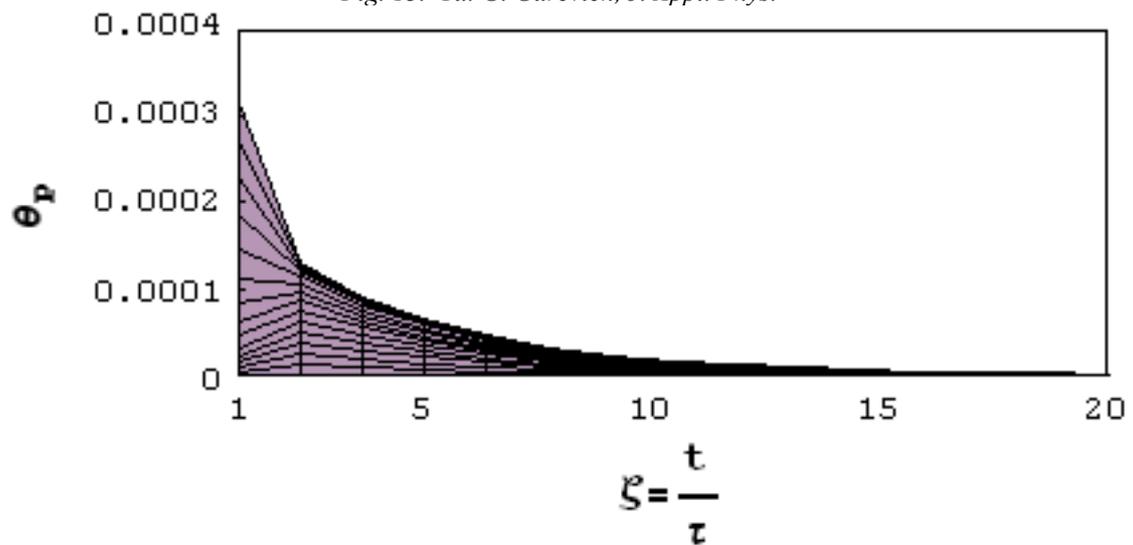

*Fig. 14. Yu. G. Gurevich, J. Appl. Phys*

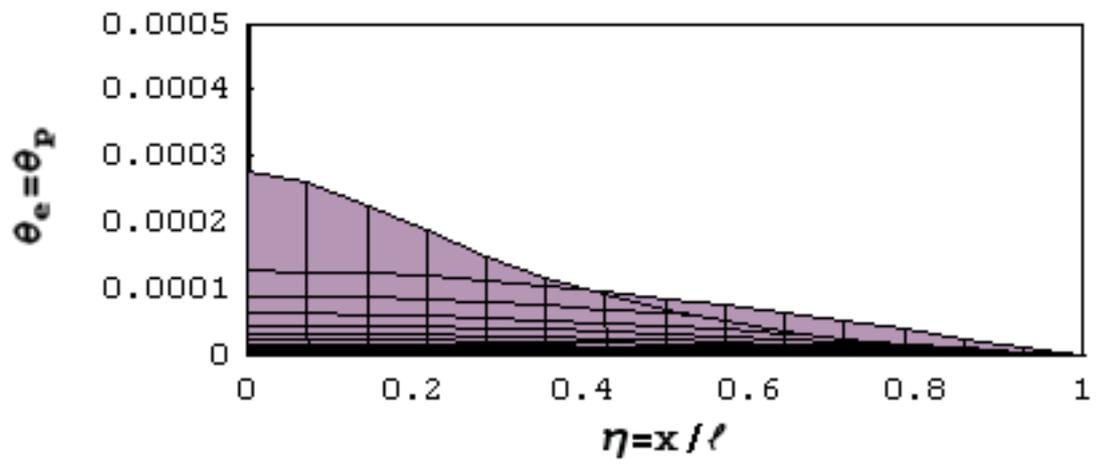

Fig. 15.  *Yu. G. Gurevich, J. Appl. Phys.*

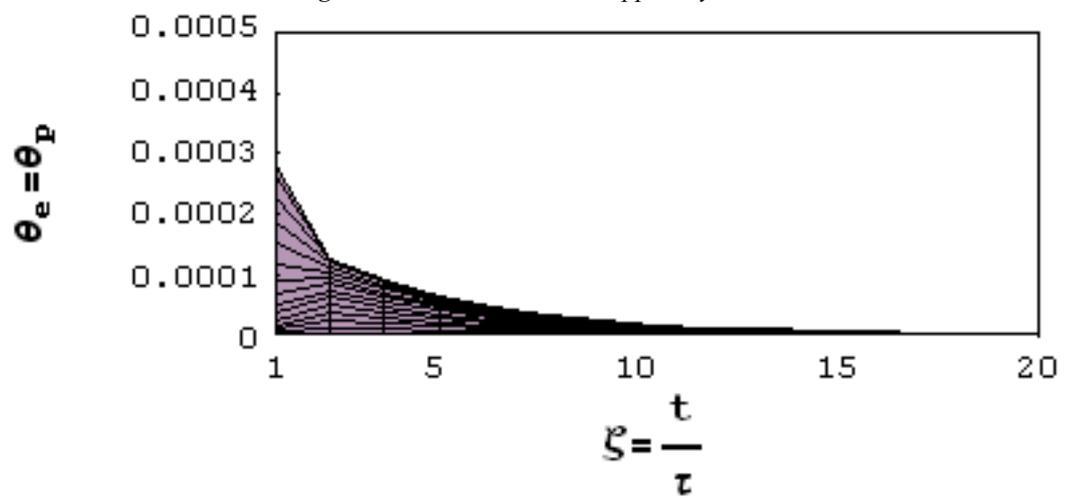

Fig. 16.  *Yu. G. Gurevich, J. Appl. Phys.*